\documentclass{article}

\usepackage{arxiv}

\usepackage[utf8]{inputenc} 
\usepackage[T1]{fontenc}    
\usepackage{hyperref}       
\usepackage{url}            
\usepackage{booktabs}       
\usepackage{amsfonts}       
\usepackage{nicefrac}       
\usepackage{microtype}      
\usepackage{lipsum}		
\usepackage{graphicx}   
\usepackage{natbib}
\usepackage{doi}
\usepackage{amsmath}
\usepackage{amssymb}
\usepackage{xcolor}
\usepackage{multicol}

\title{Diffusion assisted image reconstruction in optoacoustic tomography}

\author{ \href{https://orcid.org/0000-0003-0890-0516}{\includegraphics[scale=0.06]{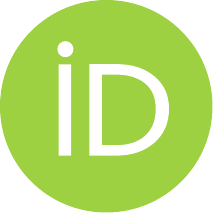}\hspace{1mm}Mart\'in G.~Gonz\'alez}\thanks{Corresponding author. Paper accepted for publication in the journal Optics and Lasers in Engineering.} \\
	Universidad de Buenos Aires and CONICET\\
	Facultad de Ingenier\'ia\\
	Buenos Aires, Argentina \\
	\texttt{mggonza@fi.uba.ar} \\
	\And
	\href{https://orcid.org/0000-0001-9180-7595}{\includegraphics[scale=0.06]{orcid.pdf}\hspace{1mm}Matias Vera} \\
	Universidad de Buenos Aires and CONICET\\
	Facultad de Ingenier\'ia\\
	Buenos Aires, Argentina \\
	\texttt{mvera@fi.uba.ar} \\
        \And
        \href{https://orcid.org/0009-0003-7943-2489}{\includegraphics[scale=0.06]{orcid.pdf}\hspace{1mm} Alan Dreszman} \\
	Universidad de Buenos Aires\\
	Facultad de Ingenier\'ia\\
	Buenos Aires, Argentina \\
	\texttt{adreszman@fi.uba.ar} \\
    \And
	\href{https://orcid.org/0000-0002-5578-0521}{\includegraphics[scale=0.06]{orcid.pdf}\hspace{1mm}Leonardo J. ~Rey Vega} \\
	Universidad de Buenos Aires and CONICET\\
	Facultad de Ingenier\'ia\\
	Buenos Aires, Argentina \\
	\texttt{lrey@fi.uba.ar} \\
}

\date{}

\renewcommand{\shorttitle}{}

\hypersetup{
pdftitle={Diffusion assisted image reconstruction in optoacoustic tomography},
pdfsubject={eess.IV, cs.AI},
pdfauthor={Martin G.~Gonzalez, Matias A.~Vera, Alan Dreszman, Leonardo J. ~Rey Vega},
pdfkeywords={Tomography, Optoacoustic, Deep Learning, Diffusion Model},
}

\begin{document}
\maketitle

\begin{abstract}
	In this paper we consider the problem of acoustic inversion in the context of the optoacoustic tomography image reconstruction problem. By leveraging the ability of the recently proposed diffusion models for image generative tasks among others, we devise an image reconstruction architecture based on a conditional diffusion process. The scheme makes use of an initial image reconstruction, which is preprocessed by an autoencoder to generate an adequate representation. This representation is used as conditional information in a generative diffusion process. Although the computational requirements for training and implementing the architecture are not low, several design choices discussed in the work were made to keep them manageable. Numerical results show that the conditional information allows to properly bias the parameters of the diffusion model to improve the quality of the initial reconstructed image, eliminating artifacts or even reconstructing finer details of the ground-truth image that are not recoverable by the initial image reconstruction method. We also tested the proposal under experimental conditions and the obtained results were in line with those corresponding to the numerical simulations. Improvements in image quality up to $17\, \%$ in terms of peak signal-to-noise ratio were observed.
\end{abstract}

%
%
%
%


\section{Introduction}
\label{sec:intro}

Optoacoustic tomography (OAT) imaging is based on the use of a laser source excitation and ultrasonic sensors in order to provide high contrast images of biological tissues maintaining at the same time great resolution \cite{Kruger_Liu_Fang_Appledorn_1995}, \cite{xu2006}, \cite{rosenthal2013}. By illuminating a biological sample with a pulsed laser, thermoelastic expansion of the sample gives rise to acoustic pressure waves that propagate through the sample. These waves can be captured by wideband ultrasonic detectors, which are placed on a detection curve/surface around the sample \cite{Minghua_Xu_Wang_2002}, \cite{Minghua_2003}, \cite{paltauf2017}. These signals, collectively known as the \emph{sinogram}, are sampled and fed into specialized numerical algorithms to reconstruct the original pressure distribution due to the laser light absorption by the illuminated sample. This is known as the \emph{acoustic inversion problem} \cite{rosenthal2013}, \cite{Hauptmann_Cox_2020}. After this initial distribution pressure problem is obtained, the so-called \emph{optical inversion problem} is solved in order to obtain a map of the optical absorption in the tissue. This is the ultimate technical goal of the OAT, as optical absorption is usually linked with important properties, among others oxygen saturation and hemoglobin concentration, that have important diagnostic value \cite{Laufer_Delpy_Elwell_Beard_2006}, \cite{Hauptmann_Cox_2020}. Although both problems are important and challenging for a full-working OAT imaging equipment, the acoustic inverse problem will be the main focus of this paper. 

The acoustic inversion problem for OAT is a well-studied subject. Depending on the details of the optical-acoustic system used for the generation of the laser illumination and the detection system used for acquiring the sinogram, the inversion can be ill-conditioned. It is for this reason that the design of robust algorithms, capable of correctly processing the detected signal and delivering a reliable estimate of the initial pressure profile, can be a challenging endeavor \cite{lutzwieler2013}. There are several approaches for the reconstruction of the pressure distribution induced by laser illumination, that can be classified as analytical or algebraic \cite{rosenthal2013}. Analytic reconstruction techniques, such as the Back-Projection (BP) \cite{xu2005} algorithms, allow the exact inversion of the forward acoustic operator by means of the analytical inversion of the mathematical equations. For example, there exist well-known analytic results for usual geometries \cite{Minghua_Xu_Wang_2002}, \cite{Minghua_2003} and that includes other effects as form factors of the acoustic sensors employed \cite{Burgholzer_Bauer-Marschallinger_Gruen_Haltmeier_Paltauf_2007}. However, analytic techniques do not take into account effects such as the presence of noise and modelling mismatches. Algebraic reconstruction techniques, on the other hand, consider the discretization or approximation of the underlying physical model (direct or inverse). Algebraic approaches, such as the model-based-matrix (MBM) algorithm \cite{Rosenthal_Razansky_Ntziachristos_2010} are well-studied and usually used as baseline benchmarks against which new reconstruction methods techniques can be compared. In order to cope with the ill-conditioned nature of the inversion problem, the use of regularization is common.  For example, it is common to include Tikhonov regularization and positivity constraints \cite{Ding_2015}, total variation constraints \cite{Huang_Wang_Nie_Wang_Anastasio_2013} or $L_1$ regularization terms that promote sparsity features in the reconstructed images or in the sinogram  \cite{Provost_Lesage_2009}, \cite{Haltmeier_Sandbichler_Berer_Bauer-Marschallinger_Burgholzer_Nguyen_2018}, \cite{Betcke_Cox_Huynh_Zhang_Beard_Arridge_2017}.

Most of the above methods are \emph{model-guided} approaches to the inverse acoustic problem, that begin with a model for the physics of the direct or forward problem. In the last few years, \emph{data-driven} approaches, where one can use measured and/or synthetic input-output pairs, gained significant traction, because of the impressive revolution brought to us by Deep Learning  \cite{LeCun_Bengio_Hinton_2015}. Deep neural nets are able to efficiently use the information contained in a dataset of numerically simulated measurements and/or true experimental data in order to find the optimal weights that can learn those aspects not contained in the ideal physical model. Although the training procedure is computationally heavy, the results are excellent, and in many cases superior to the ones obtained with the standard inversion techniques \cite{Hauptmann_Cox_2020}. Several ideas have been exploited, ranging from fully-dense convolutional nets (FD-UNets) that are able to correct artifacts introduced by the usual inversion techniques \cite{guan2020}, \cite{Guan_Khan_Sikdar_Chitnis_2020}, \cite{Awasthi_Jain_Kalva_Pramanik_Yalavarthy_2020}, to \emph{model-based} formulations \cite{Hauptmann_Lucka_Betcke_Huynh_Adler_Cox_Beard_Ourselin_Arridge_2018}, \cite{Shlezinger_Whang_Eldar_Dimakis_2020} and variational Bayesian approaches \cite{Goh_Sheriffdeen_Wittmer_Bui-Thanh_2022}, \cite{Sahlstroem_Tarvainen_2023}. 

Among variational deep learning techniques, in recent years, the so-called Diffusion Probabilistic Models \cite{Ho_Jain_Abbeel_2020} have received significant attention from the machine learning community for the task of high-resolution image generation. Inspired by non-equilibrium thermodynamics \cite{Sohl-Dickstein_Weiss_Maheswaranathan_Ganguli_2015}, these structures can sample high-quality images by learning to reverse a forward diffusion process that incrementally add noise to the original images. In this way, after the model is trained, high-quality images of the same type as the ones used to adjust the weights of the structure, can be easily generated with a far superior quality than other state-of-the-art methods \cite{Dhariwal_Nichol_2021}. These structures can also be easily adapted to construct conditional models, where the reverse process can also use conditional information like image labels information, text prompts, or a low resolution image for super-resolution image generation \cite{ho2022,Rombach_Blattmann_Lorenz_Esser_Ommer_2022}. The use of diffusion models for inverse problems from distinct fields has also been recently explored in several works \cite{Cui_Cao_Cheng_Jia_Zheng_Liang_Zhu_2023,Aali_Arvinte_Kumar_Tamir_2023, Feng_Smith_Rubinstein_Chang_Bouman_Freeman_2023,Song_Shen_Xing_Ermon_2022}.

In the present work, we will explore the use of a diffusion architecture to improve image quality reconstruction of a first initial reconstruction. Our proposal will include an initial state in which the sinogram is converted into a first reconstructed image. Then, this reconstruction is used as conditional information in a generative diffusion model to improve the quality of the image and get back, if possible, details of the ground-truth image that are not recovered in the first reconstruction. We present a discussion of the major blocks of our proposal and full assessment of its merits through numerical simulations and experimental results. The paper is organized as follows. In Section \ref{sec:reconstruction} we summarize the major mathematical details of the acoustic reconstruction problem in OAT and the basic details of diffusion models. In Section \ref{sec:method} we detail our proposal. In Section \ref{sec:results} the merits of our proposal are evaluated numerically and experimentally. Finally, in Section \ref{sec:conclu}, some concluding remarks are presented.   

\section{Models and general methods}
\label{sec:reconstruction}

\subsection{The forward and inverse problems in OAT}
\label{subsec:met_forward}

It is well-known that  after the excitation of a biological sample by an electromagnetic pulse $\delta(t)$, the acoustic pressure $p(\mathbf{r},t)$ at position $\mathbf{r} \in\mathbb{R}^3$ and time $t$, satisfies \cite{Wang_Wu_2007}:

\begin{equation}
\left(\frac{\partial^2}{\partial t^2} - v_s^2 \, \nabla^2 \right) p(\mathbf{r},t) = 0
\label{eq:waveeq}
\end{equation}

\noindent with the initial conditions,

\begin{equation}
p(\mathbf{r},0) = p_0(\mathbf{r})\, \text{,} \quad \left(\partial p /\partial t\right)(\mathbf{r},0)= 0 
\label{eq:waveeq_cond_ini}
\end{equation}

\noindent where $p_0(\mathbf{r})$ is the initial OA pressure and $v_s$ represents the speed of sound in the medium, which is assumed acoustically non-absorbing and homogeneous. Under the usual hypothesis of thermal and acoustic confinement \cite{Kruger_Liu_Fang_Appledorn_1995}, that is, when the laser pulse duration is short enough such that the heat conduction and acoustic propagation into neighboring regions of the illuminated region can be neglected, the initially induced pressure $p_0(\mathbf{r})$ is proportional to the total absorbed optical energy density. Using Green's function formalism, the pressure received by an ideal point-detector at position $\mathbf{r_d}$ can be written as:

\begin{equation}
p_d(\mathbf{r_d},t)=\frac{1}{4\pi\,v_s^2} \frac{\partial}{\partial t}\iiint_{V} \, p_0(\mathbf{r}) \frac{\delta\left(t-|\mathbf{r_d}-\mathbf{r}|/v_s\right)}{|\mathbf{r_d}-\mathbf{r}|} d^3\mathbf{r}
\label{eq:fo_time}
\end{equation}

The goal of the OAT inverse problem is to reconstruct $p_0(\mathbf{r})$ from the sinogram $p_d(\mathbf{r_d},t)$ measured at various positions $\mathbf{r_d}$, which are typically in a surface $S$ that contains the volume of interest \cite{lutzwieler2013}.

\begin{figure}
  \centering
   \hspace*{2.05cm}
  \includegraphics[width=7cm]{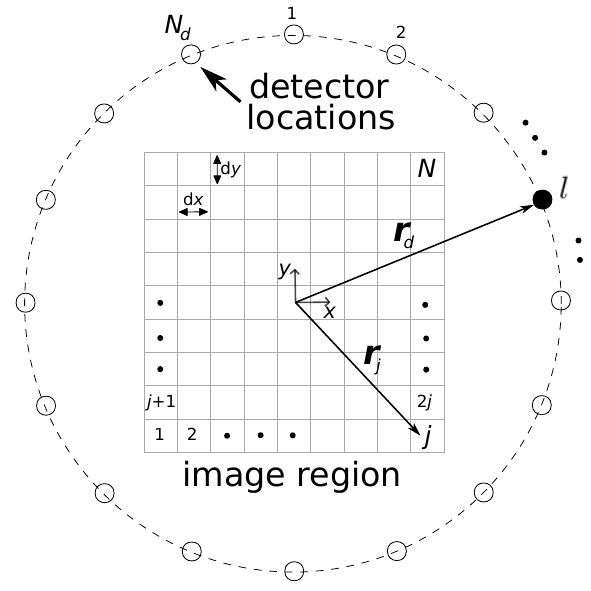}
    \caption{Schematic of the OAT imaging setup studied in this work. A number of $N_d$ detector locations are uniformly distributed around the sample, which is divided in an imaging grid of $N$ pixels.}
\label{fig:setup}
\end{figure}

Several approaches such as backprojection algorithms \cite{xu2005}, \cite{Burgholzer_Bauer-Marschallinger_Gruen_Haltmeier_Paltauf_2007} are among the most popular and used in the problem of image reconstruction in OAT. Such methods provide closed form reconstruction formulas in terms of the detected signals over the detection surface. However, it assumes that the detectors are point one with no bandwidth limitations and isotropic angular response \cite{rosenthal2013}. However, in practice, the transducers are extended, have a limited bandwidth and their spatial response is not constant.  Moreover, the detected signals are noisy. These deviations from the ideal scenario assumed by the exact reconstruction formulas can generate artifacts and distorted images.

A different but related approach to the reconstruction problem is given by a MBM algorithm \cite{Rosenthal_Razansky_Ntziachristos_2010}. In this technique, the forward solution in (\ref{eq:fo_time}) is discretized. As a result, a matrix equation is obtained which is used for solving the inverse problem. One of the advantages of this approach is that any linear effect in the system may be easily considered (e.g. sensor form factors, linear filtering or the spatial response of the sensors)\cite{Hirsch_Gonzalez_ReyVega_2021}: 

\begin{equation}
\mathbf{p_d}=\mathbf{A} \, \mathbf{p_0}
\label{eq:mbt}
\end{equation}

\noindent where $\mathbf{p_d} \in\mathbb{R}^{N_d \cdot N_t\times 1}$ is a column vector representing the measured pressures at a set of detector locations $\mathbf{r_d}_l$ ($l=1 \ldots N_d$) and time instants $t_k$ ($k=1 \ldots N_t$); $\mathbf{p_0} \in\mathbb{R}^{N\times 1}$ is a column vector representing the values of the initial acoustic pressure, and which will be typically referred as the ground-truth image; and $\mathbf{A} \in\mathbb{R}^{N_d \cdot N_t\times N}$ is the model matrix. The $j$-th element ($j=1 \ldots N$) in $\mathbf{p_0}$ contains the average value of the initial pressure within a volume element of size $\Delta V$ at position $\mathbf{r}_j$. Once the discrete formulation has been established, the inverse problem is reduced to the algebraic problem of inverting (\ref{eq:mbt}). The matrix $\mathbf{A}$ can be written as the multiplication of two matrices $\mathbf{A^{oa} \, A^s}$ where $\mathbf{A^s}$ represents the response function of the imaging system for an ideal point-like sensor and $\mathbf{A^{oa}}$ is the matrix form of a time derivative operator. The matrix $\mathbf{A^s}$ is defined as\cite{paltauf2018}: 

\begin{equation}
    A^s_{lkj} = \frac{1}{4\pi v_s^2}\frac{\Delta V}{\Delta t^2} \frac{d(t_k,\mathbf{r}_j,\mathbf{r}_{dl})}{|\mathbf{r_d}_l - \mathbf{r}_j|}
\label{eq:Gs1}
\end{equation}

\begin{equation}
    d(t_k,\mathbf{r}_j,\mathbf{r}_{dl}) = \begin{cases}
    1 & \text{si } |t_k - \frac{|\mathbf{r_d}_l - \mathbf{r}_j|}{v_s}| < \Delta t/2  \\
    0 & \text{otherwise} 
  \end{cases}
    \label{eq:Gs2}
\end{equation}

\noindent where $\Delta t$ is the time step at which the signals $p_d(\mathbf{r_d},t)$ are sampled. It is not difficult to see that (\ref{eq:Gs1}) constitutes a discretization of the integrand in (\ref{eq:fo_time}), while (\ref{eq:Gs2}) indicates the time at which the effect of initial pressure at position $\mathbf{r}_j$ is captured by the sensor $\mathbf{r_d}_{l}$.
In the case of a ﬁnite-size detector, the spatial impulse response (SIR) of the sensor is taken into account by dividing the area of the sensor into surface elements (treated as point detectors) which are then added up \cite{rosenthal2011, paltauf2018}. A typical OAT imaging setup is shown in Fig. \ref{fig:setup}.

The inversion of \eqref{eq:mbt} is typically done using a quadratic criterion plus a Tikhonov regularization term:
\begin{equation}
    \mathbf{\hat{p}_0} = \min_{\mathbf{p_0}} || \mathbf{A} \, \mathbf{p_0} - \mathbf{p_d} ||^2 + \lambda \, ||\mathbf{p_0}||^2 
    \label{eq:regTik}
\end{equation}

\noindent where $\lambda\geq 0$ is a parameter that improves the stability of the inverse problem (which is typically ill-conditioned), and it also has beneficial effects when noise is present in the measured signals. Another possibility is the use of adjoint operator \cite{Arridge_Betcke_Cox_Lucka_Treeby_2016} of $\mathbf{A}$ in order to obtain $\mathbf{\hat{p}_0}$ by:

\begin{equation}
\mathbf{\hat{p}_0}=\mathbf{A}^T\mathbf{p_d}.
\label{eq:LBP}
\end{equation}

This solution is typically known as the \emph{linear backprojection} (LBP) solution, and it is an easy and direct way to obtain a first image reconstruction. Although the image reconstructed may present several limitations and artifacts (specially in limited-view situations), it is widely used because of its simplicity. Moreover, in the context of deep learning approaches to the inverse problem in OAT, the solution of (\ref{eq:LBP}) is frequently used as the input of neural structures that are designed to improve the quality of the final reconstructed image, reducing artifacts, noise and increasing spatial resolution, \cite{Hauptmann_Cox_2020}, \cite{Gonzalez_Vera_Vega_2023}. In other works, where a neural network is used to denoise the sinogram, the LBP is used to generate the final reconstruction but from a better behaved sinogram \cite{Awasthi_Jain_Kalva_Pramanik_Yalavarthy_2020}.

\subsection{Probabilistic diffusions}
\label{subsec:diffusions}
Probabilistic diffusions are basically latent variable models that can be factorized as a Markov chain \cite{Ho_Jain_Abbeel_2020}:
\begin{equation}
p_{\theta}(\mathbf{x}_0,\dots,\mathbf{x}_T)=p_{\theta}(\mathbf{x}_T)\prod_{t=1}^Tp_\theta(\mathbf{x}_{t-1}|\mathbf{x}_t),
    \label{eq:reverse_model}
\end{equation}

\noindent where the original data $\mathbf{x}_0$ is distributed according to some unknown distribution $q(\mathbf{x}_0)$, and $\mathbf{x}_1,\dots,\mathbf{x}_T$ are intermediate latent representations, which are noisy versions (increasing noise with $t$) of $\mathbf{x}_0$. The conditional probability distributions $p_\theta(\mathbf{x}_{t-1}|\mathbf{x}_t)$ are usually modeled as multivariate Gaussian distributions $p_\theta(\mathbf{x}_{t-1}|\mathbf{x}_t)=\mathcal{N}\left(\boldsymbol\mu_t(\mathbf{x}_t,\theta),\boldsymbol\Sigma_t(\mathbf{x}_t,\theta)\right)$, where the mean and covariance matrix $\boldsymbol\mu_t(\mathbf{x}_t,\theta)$ and $\boldsymbol\Sigma_t(\mathbf{x}_t,\theta)$ are represented by deep neural networks with trainable parameters denoted by $\theta$. The distribution $p_{\theta}(\mathbf{x}_T)$ is usually kept fixed without trainable parameters and chosen as $\mathcal{N}\left(\mathbf{0},\mathbf{I}\right)$, where $\mathbf{I}$ is the identity matrix of appropriate dimensions. The full distribution $p_{\theta}(\mathbf{x}_0,\dots,\mathbf{x}_T)$ is known as the \emph{reverse process} and the marginal distribution $p_\theta(\mathbf{x}_0)$ is the ultimate-desired object (and typically impossible to be computed in closed form) that one looks to be close to the unknown true distribution $q(\mathbf{x}_0)$.  However, if we can train the full Markov chain in (\ref{eq:reverse_model}), then starting with a random $\mathbf{x}_T\sim\mathcal{N}\left(\mathbf{0},\mathbf{I}\right)$, we can go through all the steps $t=T,\dots,1$ sampling according to $p_\theta(\mathbf{x}_{t-1}|\mathbf{x}_t)$, which can be easily done because of the Gaussian model assumed for each of these distributions, and at the end keep the final generated sample $\mathbf{x}_0$. In this way, we can obtain new samples which closely match those from the unknown distribution $q(\mathbf{x}_0)$.  

The problem is how to efficiently train the model in (\ref{eq:reverse_model}). To help with this task, one can generate a \emph{forward model} as:
\begin{equation}
q(\mathbf{x}_1,\dots,\mathbf{x}_T|\mathbf{x}_0)=\prod_{t=1}^T q(\mathbf{x}_t|\mathbf{x}_{t-1})
    \label{eq:forward_model}
\end{equation}

\noindent where $q(\mathbf{x}_t|\mathbf{x}_{t-1})$ has no trainable parameters and is distributed as $\mathcal{N}(\sqrt{1-\beta_t}\mathbf{x}_{t-1},\beta_t\mathbf{I})$, with $\beta_t$ is fixed sequence of real numbers less than 1, that are known as the $\emph{noise schedule}$ \cite{Nichol_Dhariwal_2021}. The forward model starts from a known sample $\mathbf{x}_0$ from the unknown distribution $q$ and generates a sequence of increasingly noisy versions (the amount of noise controlled by $\beta_t$) $\mathbf{x}_t$ to finally delivers $\mathbf{x}_T\sim\mathcal{N}\left(\mathbf{0},\mathbf{I}\right)$, which as explained above, will be the input of the reverse model.  With this forward model, we can set up the following loss function to obtain the parameters $\theta$ of the reverse model (see the mathematical details in \cite{Ho_Jain_Abbeel_2020}):

\begin{equation}
L(\theta)=\frac{1}{N}\sum_{n=1}^N \left(\sum_{t>1}^T D_{\rm KL}(q(\mathbf{x}_{t-1}|\mathbf{x}_t,\mathbf{x}_0^{n})\,||\,p_{\theta}(\mathbf{x}_{t-1}|\mathbf{x}_t))-\log p_{\theta}(\mathbf{x}_0^n|\mathbf{x}_1)\right),
\label{eq:cost_function}
\end{equation}

\noindent where $\left\{\mathbf{x}_0^n\right\}_{n=1}^N$ is the training set (distributed according to the unknown distribution $q(\mathbf{x}_0)$), $D_{KL}(\cdot)$ is the usual Kullback-Leibler distance between probability distributions \cite{Cover_Thomas_2006}, and $q(\mathbf{x}_{t-1}|\mathbf{x}_t,\mathbf{x}_0^{n})$ can be easily obtained from (\ref{eq:forward_model}) and shown to be a Gaussian multivariable with mean depending on $\mathbf{x}_t$, $\mathbf{x}_0^n$ and $\beta_t$ and diagonal covariance matrix depending on $\beta_t$. As all distributions are Gaussian, Kullback-Leibler distances can be computed in closed form, $\log p_{\theta}(\mathbf{x}_0|\mathbf{x}_1)$ is also easily calculated and (\ref{eq:cost_function}) becomes a quadratic cost function in terms of the training set that can be optimized for $\theta$ using standard backpropagation and gradient descent techniques. The loss function is finally written as (assuming that $\boldsymbol\Sigma_t(\mathbf{x}_t,\theta)=\sigma_t^2\mathbf{I}$ with $\sigma_t$ fixed):

\begin{equation}
L(\theta)=\frac{1}{N}\sum_{n=1}^N \sum_{t=1}^T \gamma_t\|\boldsymbol{\epsilon}_t-\boldsymbol{\epsilon}_\theta(\overbrace{\sqrt{\bar{\alpha_t}}\mathbf{x}_0^n+\sqrt{1-\bar{\alpha}_t}\boldsymbol{\epsilon}_t}^{\mathbf{x}_t},t)\|^2 ,
\label{eq:cost_function_final}
\end{equation}

\noindent where $\gamma_t$ is a parameter that depends on sequences $\beta_t$ and $\sigma_t$, $\bar{\alpha_t}=\prod_{s=1}^{t}(1-\beta_s)$, $\boldsymbol{\epsilon}_t$ is the Gaussian corrupting noise sampled from $q(\mathbf{x}_t|\mathbf{x}_{t-1})$ in the forward $t$-step, and $\boldsymbol{\epsilon}_\theta(\sqrt{\bar{\alpha_t}}\mathbf{x}_0^n+\sqrt{1-\bar{\alpha}_t}\boldsymbol{\epsilon}_t,t)$, which can be obtained from $\boldsymbol\mu_t(\mathbf{x}_t,\theta)$,
is the \emph{denoising noise} in the reverse $t$-step introduced by $p_{\theta}(\mathbf{x}_{t-1}|\mathbf{x}_t)$.

The rationale behind (\ref{eq:cost_function}) and (\ref{eq:cost_function_final}) is that if the noise perturbations in each step of the forward model are carefully tuned (and typically small), the information corrupted by this noise can be recovered in the corresponding step of the reverse model. In fact, each term in  (\ref{eq:cost_function}) considers the distance between the denoising distribution $p_{\theta}(\mathbf{x}_{t-1}|\mathbf{x}_t)$ at $t$-step in the reverse model and the easily computable posterior distribution $q(\mathbf{x}_{t-1}|\mathbf{x}_t,\mathbf{x}_0^{n})$ at the $t$-step of the forward model.  Similarly, it is seen in (\ref{eq:cost_function_final}) that in the reverse $t$-step, a parametrized neural network $\boldsymbol{\epsilon}_\theta$ with input $\sqrt{\bar{\alpha_t}}\mathbf{x}_0^n+\sqrt{1-\bar{\alpha}_t}\boldsymbol{\epsilon}_t$ seeks to approximate the corrupting noise $\boldsymbol{\epsilon}_t$ sampled during the forward $t$-step.

This is the basic idea behind a diffusion-based generative model. However, in inverse problems (which is our case) we are interested in the reconstruction of $\mathbf{x_0}$ based on the measurement of sample $\mathbf{y}_0$ which are statistically related through an unknown conditional distribution $q(\mathbf{x}_0|\mathbf{y}_0)$. For example, for our application, $\mathbf{x}_0$ would represent the true image and $\mathbf{y}_0$ the measured sinogram or some initial processing of it. The idea behind using $\mathbf{y}_0$ is that, by exploiting the statistical dependence with the desired object $\mathbf{x}_0$, the diffusion process can be efficiently ``biased'' to better capture the statistical information shared by them. In order to cover this situation, we can easily modify the above criterion and consider a reverse conditional model \cite{ho2022}, \cite{Rombach_Blattmann_Lorenz_Esser_Ommer_2022} $p_{\theta}(\mathbf{x}_{t-1}|\mathbf{x}_t,\mathbf{y}_0)$ for $t=1,\dots, T$, that is also modeled by a multivariable Gaussian with trainable mean and covariance $\boldsymbol\mu_t(\mathbf{x}_t,\mathbf{y}_0,\theta)$ and $\boldsymbol\Sigma_t(\mathbf{x}_t, \mathbf{y}_0,\theta)$. The loss function can be finally written as:

\begin{equation}
L(\theta)=\frac{1}{N}\sum_{n=1}^N \sum_{t=1}^T \gamma_t\|\boldsymbol{\epsilon}_t-\boldsymbol{\epsilon}_\theta(\sqrt{\bar{\alpha_t}}\mathbf{x}_0^n+\sqrt{1-\bar{\alpha}_t}\boldsymbol{\epsilon}_t,g_\theta(\mathbf{y}_0^n),t)\|^2 ,
\label{eq:cost_function_final2}
\end{equation}

\noindent where $\left\{\mathbf{x}_0^n,\mathbf{y}_0^n\right\}_{n=1}^N$ is the training set and where $g_\theta$ is a trainable representation of the conditional information $\mathbf{y}_0$ for efficient processing in the reverse diffusion process.

\begin{figure}
  \centering
  \includegraphics[width=15cm]{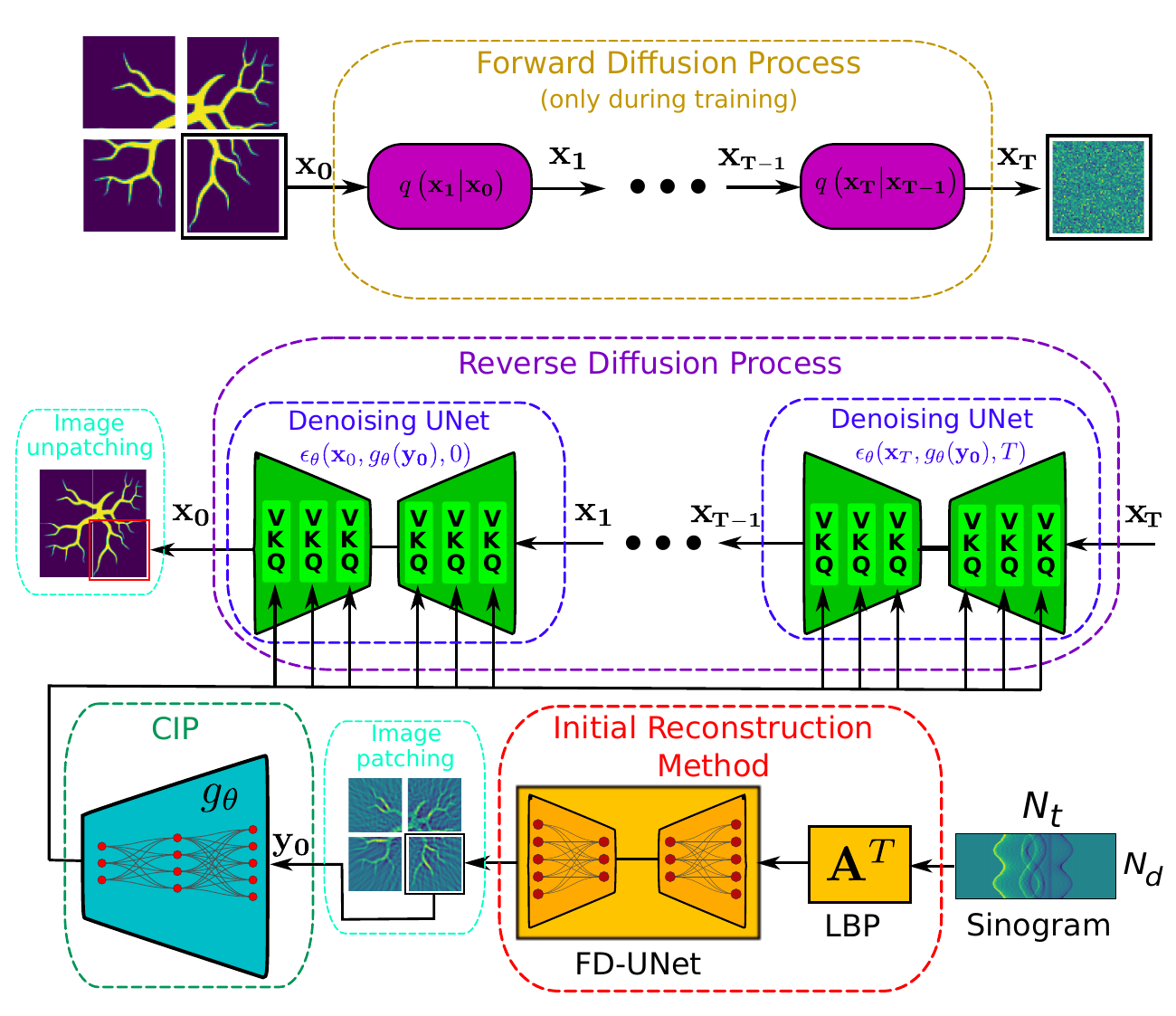}
    \caption{Proposed method architecture. The forward diffusion process only acts during training phase. $3-$tuples $(\mathbf{V},\mathbf{K},\mathbf{Q})$ indicate the multi-head cross-attention mechanisms at each UNet's scales. Sinusoidal positional time embeddings are used for representing the time-steps in the reverse diffusion process. }
\label{fig:arch}
\end{figure}

\section{Proposed method architecture}
\label{sec:method}

Our proposal is to use a conditional diffusion model to improve the image quality reconstructed with a standard and well-proven method. The main idea is to use the image reconstructed by this well-proven method as conditional information to a diffusion model that will enhance the final image and eliminate possible artifacts besides increasing resolution (if desired).  Three major blocks can be identified in the proposed method: 1) the initial well-proved reconstruction method, 2) the conditional information preprocessing (CIP), and 3) the conditional diffusion model in reduced dimension. In Fig. \ref{fig:arch} we can see a depiction of the different blocks and their interconnections.

\subsection{Initial reconstruction method}
The measured sinogram is processed by matrix $\mathbf{A}^T$ generating an initial image corresponding to the LBP solution in (\ref{eq:LBP}). This method is simple to implement and numerically efficient \cite{Hoelen_Mul_2000} but usually introduces some artifacts, specially when the number of acoustic sensors around the sample are limited and the signal-to-noise ratio (SNR) is not large enough. This solution can also suffer from possible mismatches between the nominal measurement system assumed to construct matrix $\mathbf{A}$ and the actual one (e.g. sensor position uncertainties, assumed sound velocity in the sample, etc) \cite{Hirsch_Gonzalez_ReyVega_2021,vera2023}. For this reason, this initial image is fed into an appropriate neural network, with a sufficiently large expressive capacity power, in order to correct artifacts and improve quality. This neural network is trained  separately to the other blocks in our system. In particular, we chose a UNet architecture \cite{Ronneberger_Fischer_Brox_2015}, which is basically a multi-scale convolutional autoencoder with residual connections between and skip connections between encoder and decoder at each scale. At each scale, we also include dense connections to improve information flow along the network and provide robustness. This leads to a Fully-Dense UNet (FD-UNet) architecture, which is well-known to be a competitive solution for OAT image reconstruction \cite{guan2020}. In fact, this architecture and its parameters are essentially the same to those considered in \cite{Gonzalez_Vera_Vega_2023}, where the specific details are discussed. The loss used for this FD-UNet structure was also the mean square error (MSE) between the ground-truth and reconstructed images. The only meaningful change is the fact that in this paper only one frequency band is used ($n=1$) and the number of spatial sensors is now $N_d=36$. One important issue that could influence the output reconstructed image from this stage is the fact that the forward model $\mathbf{A}$ is used in the LBP. In practical situations, the exact forward model is influenced by parameters such as the sensor positions and the speed of sound in the sample. As those parameters are not always exactly known, some nominal values are assumed for constructing matrix $\mathbf{A}$. It is well-known that mismatch between these nominal assumed values and the true ones is responsible for decreased quality of the obtained images \cite{Sahlstrom2020}. In \cite{Gonzalez_Vera_Vega_2023} and \cite{vera2023}, the robustness of a LBP+FD-UNet architecture (against uncertainties in sensor positions and speed of sound among others) was extensively studied and concluded that the said architecture is robust enough under reasonable deviations against nominal values used for constructing matrix $\mathbf{A}$. In any case, it is in this initial reconstruction stage where such effects should be taken care of, being the conditional reverse diffusion enhancement described in Section \ref{sec:conditional} almost oblivious to them. It is worth to mentioning that, although this is our choice for the initial reconstruction method which will be used for the conditional information for the diffusion model, other choices are possible. As long as the chosen algorithm has not trainable parameters (e.g. (\ref{eq:regTik})), or they are trained separately to the other blocks, other solutions can be easily considered.  

\subsection{Conditional information preprocessing (CIP)}
\label{sec:CIP}

The initial reconstructed image from the previous block will be used as conditional information for the diffusion process. This information is passed to the diffusion model through a trainable function $g_\theta$. In our case, we considered $g_\theta$ as the encoder from a vanilla autoencoder with 3 hidden layers with sizes of 3072, 2048 and 1024 and ReLu activations. This autoencoder was optimized using the well-known mean square error (MSE) loss and discarding the decoder part after training. The exact values for the hidden layer sizes were obtained from a hyperparameter tuning optimization procedure. The output of this encoder is the conditional information used in the diffusion model. 

As will be explained in the next subsection, the conditional diffusion model will not work in the original image space, but in a lower-dimensional one. Working at image level can be computational demanding and several alternatives can be used. For example, a popular choice is to consider an appropriately tuned latent space for the diffusion model. Each image can be compressed by a variational autoencoder with a Kullback-Leibler loss \cite{Kingma2014} and both forward and reverse diffusion processes work in the latent space, which is of considerably smaller dimension than the original image space. The encoder is used only in the training phase and it mainly affects the forward process. The decoder is used to return to the original image space the latent representation generated by the reverse process. It is used both at training and inference phase. This autoencoder is not necessarily jointly optimized with the diffusion model. This is for example the approach taken in \cite{Rombach_Blattmann_Lorenz_Esser_Ommer_2022}, where the results of this architecture for super-resolution image are excellent. In our case, this approach did not give good results. A possible explanation is the fact that for an encoder with a high compression rate, any small noise or artifact introduced by the reverse diffusion model working in the latent space, could generate an important non-desired effect in the image at the decoder output. It seems that a very careful design and training of the variational autoencoder for the latent space representations is needed. This could be difficult to achieve, especially if there is not a large training set of images such as in our case.

Another possibility, that it is explored in this work, is to consider a patching of the images to be processed. The images used, not only as the input for the forward diffusion process, but also for the conditional information preprocessing are patched in smaller images. Each of these patches are going through the forward diffusion model (only during training) and the same patching strategy is used for the initial reconstructed images. That is, the input to the autoencoder $g_\theta$ defined above, is patched in the same exact way and processed to obtain the conditional representation of each initial reconstructed image patch, which is then introduced into the reverse diffusion model.  This imposes that the generative process at the diffusion model will work at patch level. After all patches of a given image are generated using the corresponding conditional information, the final image is assembled.

\subsection{Conditional diffusion model in reduced dimension}
\label{sec:conditional}

The conditional diffusion model generates refined reconstructions for each image patch using the conditional information generated as explained above from initial image reconstruction and initial noisy image patch sampled as  $\mathbf{x}_T\sim\mathcal{N}\left(\mathbf{0},\mathbf{I}\right)$. During training, samples corresponding to the patches corresponding to the  ground-truth images are degraded during $T$ steps by the forward diffusion process, which has no-trainable parameters and where the main design choice is the noise schedule $\left\{\beta_t\right\}_{t=1}^T$. This sequence is a typically linear sequence starting from $\beta_1$ which is typically a small value (e.g. $10^{-4 }$) and ending in a larger $\beta_T$ (e.g. 0.02). The remaining choice of $T$ is critical. A large $T$ will guarantee a more fine-grained control of the noise corruption of the ground-truth image patches by the forward process (during the training phase) and that the Gaussian assumption for  $p_{\theta}(\mathbf{x}_{t-1}|\mathbf{x}_t,g_\theta(\mathbf{y}_0))$ for $t=1,\dots, T$ is a good approximation for the true reverse process and where $\mathbf{y}_0$ is the initial image reconstruction. However, this presents the problem that, during inference, the same exact number of reverse steps has to be implemented to obtain the final reconstructed image patches through the sequential sampling. As this has to be done sequentially and not in parallel, the process of generating the final reconstructed high quality patches is a very slow task. In order to overcome these shortcomings, we followed the approach in \cite{Song_Meng_Ermon_2022}, where some straightforward modifications of the sampling procedure of $p_{\theta}(\mathbf{x}_{t-1}|\mathbf{x}_t,g_\theta(\mathbf{y}_0))$ for $t=1,\dots, T$ during inference are proposed for significantly reducing the computational burden of that task. The main idea consists in finding a non-Markovian equivalent forward process (that leads to the same loss function considered in (\ref{eq:cost_function_final2}) and to the exactly same trained neural structures). Interestingly enough, choosing carefully the non-Markovian dynamics, one can obtain a Markovian reverse diffusion process with a significantly smaller number of steps which speed up the inference stage. Further details can be found in the above mentioned reference \footnote{See also \url{https://huggingface.co/docs/diffusers/api/schedulers/ddim} for implementation details.}.

The conditional information corresponding to the initial image reconstruction patches is introduced in the reverse diffusion process as explained above (i.e. $p_{\theta}(\mathbf{x}_{t-1}|\mathbf{x}_t,g_\theta(\mathbf{y}_0))$ ). The cost function given by (\ref{eq:cost_function_final2}) is then optimized using backpropagation techniques. Each step $t=1,\dots, T$ in the reverse process (i.e. $\epsilon_\theta(\mathbf{x}_t,g_\theta(\mathbf{y_0}),t)$) is modeled by a UNet architecture\footnote{We considered the default parameters implemented in \url{https://huggingface.co/docs/diffusers/api/models/unet2d-cond}. We only modified the number of output channels at each scale with respect to the default ones to the following values (the same for each time-step $t$): (128, 256, 512, 1024). This was done for optimizing the use of GPU memory during training. Also, as we are not working with RGB images, the number of channels at the input and output for the UNet were set to 1.}. Notice that in principle, the figure indicates that for each time-step there is a different UNet. This is taken care of with input $t$ in $\epsilon_\theta(\mathbf{x}_t,g_\theta(\mathbf{y_0}),t)$ which is implemented through a sinusoidal time embedding as proposed in \cite{Ho_Jain_Abbeel_2020}. In this way, parameters can be shared across time-steps. At each scale, a number of ResNet layers \cite{He_Zhang_Ren_Sun_2016} are implemented to improve parameter training. The different scales of the UNet also include the conditional information $g_\theta(\mathbf{y_0})$ (both at the encoder and the decoder) through the use of a multi-head cross-attention structure \cite{Vaswani_Shazeer_Parmar_Uszkoreit_Jones_Gomez_Kaiser_Polosukhin_2017} which is jointly trained with the ResNet layers and the conditional information preprocessing $g_\theta(\cdot)$. 

In summary, our proposal uses several well-known architectures (e.g., autoencoders, UNets, attention mechanisms, diffusions, etc) which have been extremely successful in diverse machine learning tasks. However, they are specifically tuned for our application of OAT image reconstruction, not only to improve performance specifically for this application, but also to reduce the computational burden required to train the models and produce inferences with them. In fact, as explained in the next section, after training is completed, the proposed method is computationally competitive with other standard reconstruction methods in OAT.

\section{Numerical and experimental results}
\label{sec:results}

\subsection{Implementation details}
\label{subsec:details}

To test the proposed method, we used a setting similar to the one in Fig. \ref{fig:setup} with images of $128\times 128$ pixels (pixel size of $110\, \mu\text{m}$) and $N_d=36$ detector locations uniformly distributed around the sample. The distance from the center of the sample to the circle where the sensor is positioned ($\mathbf{r}_{d}$) is $44 \text{ mm}$. The sampling frequency was set to $24.4 \text{ MHz}$. The detector locations had a random uncertainty of position of $0.1 \%$.  Each pressure signal had a duration of $25 \,\mu\text{s}$ ($N_t=1024$ samples). In order to generate the sinograms corresponding to the ground-truth images needed to train the proposed architecture we made use of the program j-Wave\footnote{\url{https://ucl-bug.github.io/jwave/index.html}} \cite{Stanziola_Arridge_Cox_Treeby_2023}, which is a highly-customizable acoustics simulator based on Python and JAX. Using this tool we followed a simple approach to construct the matrix forward operator. For each of the $N$ pixels of the images we considered a delta pressure function centered at such a pixel. Then, with the help of the j-Wave, we obtained the response of such pressure function at each of the sensor locations. Such a response is a vector of dimension $N_d\times N_t$. This vector is technically a column of the forward operator (the one corresponding to the activated pixel). Repeating this for each pixel, we are able to numerically construct the full operator. When the forward operator is constructed in this way, we can generate the sinogram at the sensor locations for an arbitrary image (assuming linearity of the forward operator) by multiplying the original image by the matrix forward operator. The results of using this approach for generating the sinogram are almost identical to performing a full j-Wave simulation for each image. In this way, the time of constructing each sinogram is significantly shorter than performing a full simulation with j-Wave. On top of these generated sinograms white Gaussian noise was added with variable variance levels, leading to sinogram measurement SNR values between $20 \text{ dB}$ and $80 \text{ dB}$. For constructing the matrix $\mathbf{A}$, which will be used for the implementation of the LBP solution, we considered a speed of sound  of $v_s = 1490 \text{ m/s}$ and used equations (\ref{eq:Gs1}) and (\ref{eq:Gs2}). We also included, in matrix $\mathbf{A}$, the impulsive and spatial response of a widely used commercial ultrasonic sensor (Olympus V306-SU)\cite{reigada2023}, which have a circular active area with a diameter of $13 \text{ mm}$.

We found sufficient to work with 4 non-overlapping patches for each original image (that is, each patch has a size of $64 \times 64$). The conditional information preprocessing function $g_\theta$ is the encoder of the autoencoder with 3 hidden layers, whose input dimension is the one corresponding to the image patches and its output is a vector of length $1024$ as was explained in Section \ref{sec:CIP}. This vector is the conditional information introduced in the reverse diffusion process that generates the improved image patches, which are then used to construct the full image with size of $128 \times 128$. The number of $T$ for the diffusion model was set to $1000$ and the noise schedule start at $\beta_1=10^{-4}$ and linearly increased to $\beta_T=0.02$.

For training, we used a dataset containing synthetic and experimental retinal vasculature phantoms from public databases \cite{drive2020,aria2006,rite2013,stare2000,hatamizadeh2022ravir}. The number of phantoms used for training were $64,000$ and $10,000$ were used for validation.  We used ADAM optimization \cite{Kingma2015AdamAM} with parameters $\alpha_1=0.9$ and $\alpha_2=0.999$. The initial learning rate, number of epochs and the batch size were set to $10^{-4}$, 200 and 16, respectively. The total training time was approximately 5 days in a computer with CPU Intel i9-10900F, 128 GB of RAM and two GPU RTX  3090 each with 24 GB of memory. 

\subsection{Numerical results}

Table \ref{table:1} shows the results of our proposal at inference after the full training of the models. We computed the average of several metrics over $600$ ground-truth images not present in the training data set. The metrics considered in this work are the Structural Similarity Index (SSIM) and the Peak Signal to Noise Ratio (PSNR), which are popular metrics to evaluate the performance of image reconstruction algorithms. The entry corresponding to our proposal is DAR (Diffusion Assisted Reconstruction). In the first and second rows, we considered the metrics for DAR with a number of inference steps (NIS) of 5 and 25, respectively. This number refers to the number of inference steps used during the inference phase in the reverse diffusion model in line with the approach suggested in \cite{Song_Meng_Ermon_2022} as explained in Section \ref{sec:conditional}.  We also included, for the sake of comparison, the performance of both the LBP and the FD-UNet methods which are part of the initial reconstruction method of our proposal. In this way, we are efficiently evaluating the performance gains of our proposal with respect to the baselines corresponding to those methods.

\begin{figure}
  \centering
  \includegraphics[width=17cm]{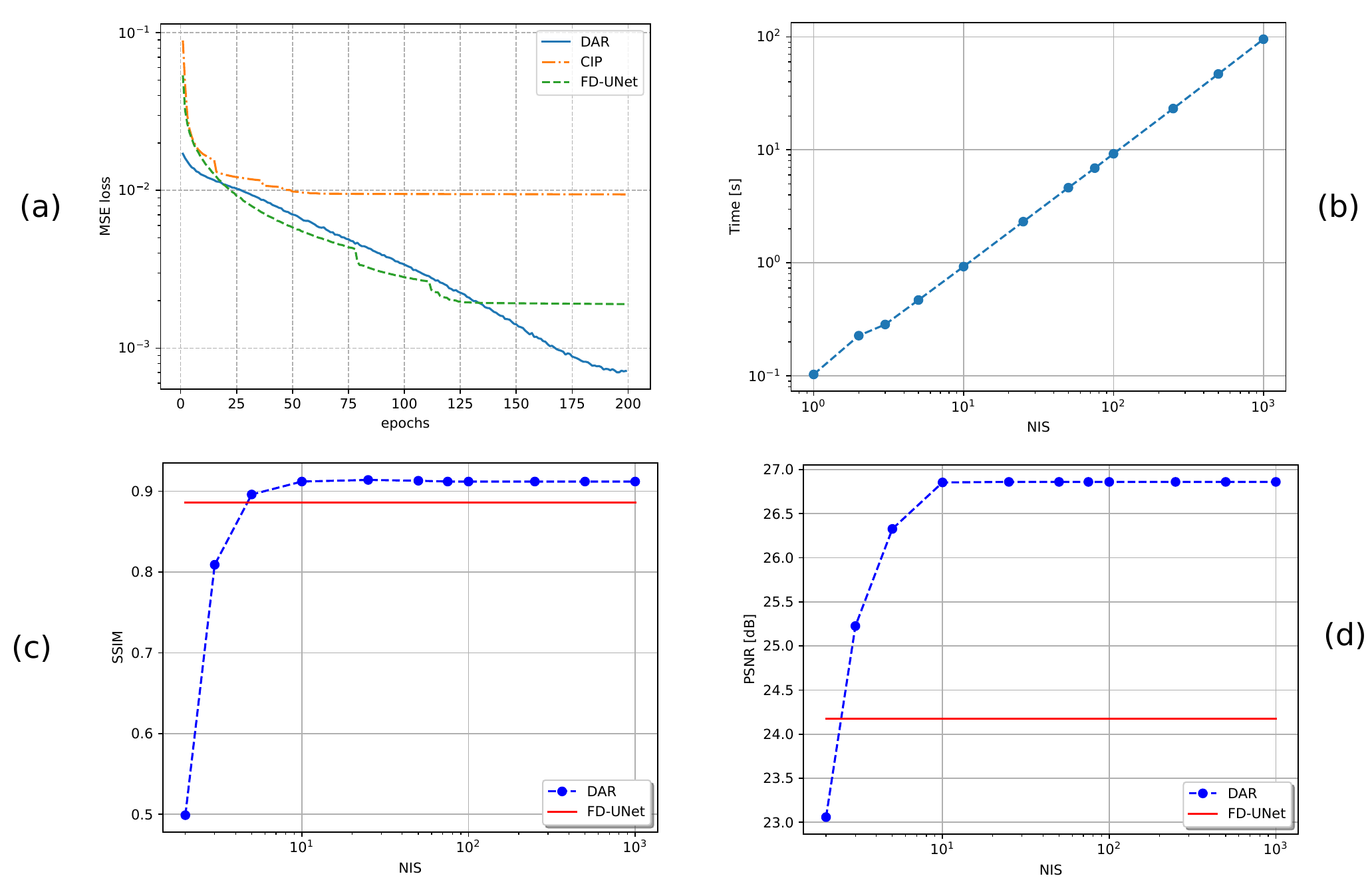}
    \caption{(a) Evolution of the losses with the number of epochs during training for the conditional diffusion model and the FD-UNet and CIP blocks. (b) Time required for the DAR to make a reconstruction of a full image of $128\times128$ partitioned in $4$ patches of $64\times64$ as a function of the number of inference steps. (c) Peak Signal to Noise Ratio as a function of the number of inference steps. (d) Structural Similarity Index as a function of the number of inference steps.}
\label{fig:4fig}
\end{figure}

In Fig. \ref{fig:4fig}.a we present the training evolution of the losses of the three major blocks. The diffusion model is trained according to (\ref{eq:cost_function_final2}), which is basically a MSE loss because of the Gaussian assumptions for the latent variable models (as mentioned in Section \ref{subsec:diffusions}). The FD-UNet and CIP blocks were trained using the MSE loss between the reconstructed images and the ground-truth ones, as explained in Section \ref{sec:method}. In Fig. \ref{fig:4fig}.a can be appreciated that the training loss for the FD-UNet and CIP converge rapidly to the steady state, while the diffusion model takes more time in reaching it. However, the steady state of the diffusion model is lower than the one corresponding to the FD-UNet, which is also indicative of the superior performance of the diffusion model.

To better analyze the effect of the influence of the level of measurement noise, in Table \ref{table:2}, we present the performance results for varying levels in the measurement SNR for both SSIM and PSNR metrics. Unsurprisingly, we see that in general the performance improves with increasing values of SNR. The SSIM metric is less sensitive to changes in the SNR (at least in the levels considered) than the PSNR metric. It is interesting to observe that the performance of the LBP under both performance metrics is insensitive to the level of SNR. 
This can be explained by the fact that, for this method, performance is primarily influenced by the artifacts and defects introduced by the method itself, which appear to be of greater magnitude than the detrimental effects that noise can cause.

It is clear that, although the results corresponding to the LBP are very poor, the use of an FD-UNet over the reconstructed images delivered by that method, provides significant gains. Similarly and as expected, our proposal has an extra gain with respect to the performance of the FD-UNet. This gain is systematic for the two metrics considered. Moreover, the performance gains even happen for an extremely low number of NIS ($5$) and present further improvements when the number of NIS increases.  
As the computational complexity at inference is dominated by the NIS, it is important to quantify the trade-offs between performance and complexity. Figs. \ref{fig:4fig}.c and \ref{fig:4fig}.d show the performance in terms of SSIM and PSNR for the DAR with a varying number of NIS (up to $1000$ steps), respectively. The performance of the FD-UNet is also displayed for easy reference (although it remains obviously constant as NIS increases). We see that, increasing NIS beyond some point (typically 25), the performance gain of our proposal is negligible. It is also seen that a number of 5 NIS is already providing performance gains with respect to the FD-UNet output. This is very satisfying from the practical point of view. In Fig. \ref{fig:4fig}.b, we see the time required to perform a full reconstruction of an image of size of $128\times 128$ as a function of the NIS. As expected, the relation is almost linear. We see that an image reconstruction with NIS below $100$ steps takes almost a second, which is negligible with respect to the time required to collect the sinogram and turns it into a reasonable number of steps for the application. 

\begin{table}[t]
    \centering \caption{Performance (mean value and standard deviation) over the testing set.}
    \begin{tabular}{ccc}
        \hline
        Method & SSIM & PSNR \\
        \hline
        DAR 5 & 0.896 $\pm$ 0.079 & 26.326 $\pm$ 6.468 \\
        DAR 25 & \textbf{0.914} $\pm$ 0.076 & \textbf{26.859} $\pm$ 7.570 \\
        FD-UNet  & 0.886 $\pm$ 0.073 & 24.176 $\pm$ 3.449 \\
        LBP  & 0.182 $\pm$ 0.040 & 8.764 $\pm$ 0.76 \\
        \hline
   \end{tabular}
\label{table:1}
\end{table}

\begin{table}[t]
    \centering \caption{Average performance over the testing set as function of SNR (for SSIM and PSNR).}
    \begin{tabular}{cccc}
    \multicolumn{4}{c}{SSIM} \\
        \hline
        Method & 20 dB & 40 dB & 60 dB \\
        \hline
        DAR 5 & 0.886 & 0.896 & 0.896 \\
        DAR 25 & \textbf{0.906} & \textbf{0.915} & \textbf{0.915}  \\
        FD-UNet  & 0.869 & 0.887 & 0.888 \\
        LBP  & 0.182 & 0.182 & 0.182 \\
        \hline   
   \end{tabular}
 \hspace{0.6cm}     
       \begin{tabular}{cccc}
       \multicolumn{4}{c}{PSNR} \\
        \hline
        Method & 20 dB & 40 dB & 60 dB \\
        \hline
        DAR 5 & 25.486 & 26.438 &  26.462 \\
        DAR 25 & \textbf{25.857}  & \textbf{26.964} & \textbf{27.021}  \\
        FD-UNet  & 23.688  & 24.240 &  24.272 \\
        LBP  & 8.760  & 8.762 & 8.763 \\
        \hline
   \end{tabular}
\label{table:2}
\end{table}

\begin{figure}
  \centering
  \includegraphics[width=16cm]{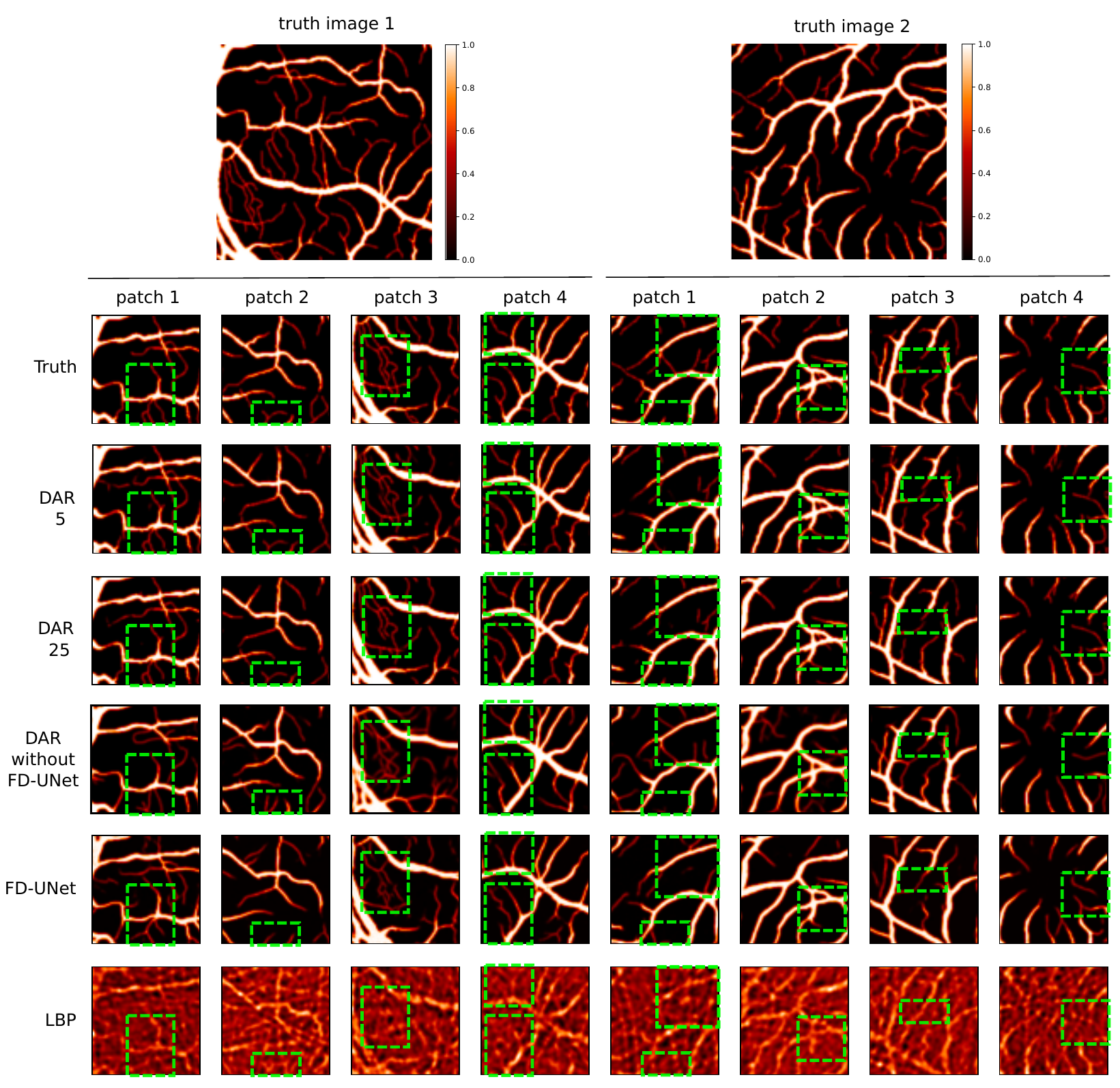}
    \caption{Qualitative examples for DAR (NIS=5 and 25), DAR without the FD-UNet block (NIS = 25), FD-UNet and LBP. Green boxes indicate finer details captured by DAR with increasing NIS and with respect to FD-UNet reconstructions.}
\label{fig:simulres}
\end{figure}

In Fig. \ref{fig:simulres} we present some qualitative results that complement the average results presented in Table \ref{table:1}. We consider the results for two different images of blood vessels in the testing set (not present in training) for all the methods considered. The DAR results are for two different NIS: $5$ and $25$. For easier visualization, we separately present the qualitative results for each of the four patches for each image. We added green boxes in each patch in order to indicate finer details (e.g. areas with numerous small vessels), that are better reconstructed by the DAR with respect to FD-UNet and/or different value of NIS. For example we see that in the patch one of image 1, the FD-UNet does not reconstruct with sufficient details the morphology of the small vessels at the bottom. The DAR is able to recover with better quality those details, being more similar to the ones present in the ground-truth image. Patch 4 of image 2 is another clear example where the FD-UNet is not able to recover some small vessels, while they are recovered by DAR for both values of NIS. Other green boxes focus on details that are present on the ground-truth images that are recovered by DAR with varying degrees of quality depending on the number of NIS. For example, in patch 1 of image 2, we see that DAR with NIS 5 present a ``hallucination'' including a small vessel that is not present in the ground-truth image. However, DAR with NIS 25 is in close match with the ground-truth image not presenting that erroneous vessel. Patch 3 of image 1 is another example, where DAR with NIS 25 is able to recover finer details of a group of small vessels that is present in the ground-truth image. 

\begin{table}[t]
    \centering \caption{Performance (mean value and standard deviation) over the testing set for DAR using only LBP as initial reconstruction method.}
    \begin{tabular}{ccc}
        \hline
        Method & SSIM & PSNR \\
        \hline
        DAR 5 & 0.866 $\pm$ 0.089 & 23.078 $\pm$ 7.451 \\
        DAR 25 & \textbf{0.889} $\pm$ 0.077 & \textbf{23.648} $\pm$ 7.066 \\
        LBP  & 0.182 $\pm$ 0.040 & 8.764 $\pm$ 0.76 \\
        \hline
   \end{tabular}
\label{table:3}
\end{table}

Our proposal for the initial reconstruction method included the LBP+FD-UNet. The use of an FD-UNet is guided primarily by the fact that previous works show that it is a robust method for OAT image reconstruction. However, there is not the only possibility that could be considered. The initial reconstruction method only set the baseline from which the diffusion processing could generate improvements and enhancements in image quality. In order to test this, we considered an alternative architecture in which the initial reconstruction method only included a basic LBP reconstruction. We trained the CIP and the diffusion model considering this. The results over the 600 ground-truth images used for inference are summarized in Table \ref{table:3} where, for easier reference, we also show the LBP performance. It can be easily appreciated that the DAR provides a gain (a significant one) with respect to the LBP. In comparison with Table \ref{table:1}, we can also see that the performance (for both metrics, SSIM and PSNR) of DAR with only LBP as initial reconstruction method is inferior to the case of DAR with LBP+FD-UNet as initial reconstruction method, showing effectively that the initial reconstruction method set the baseline performance from which further gains can be expected. The fourth row in Fig. \ref{fig:simulres} shows the qualitative results DAR without FD-UNet with a value of NIS equal to 25. It also easily observed that the scheme provides a significant enhancement with respect to the LBP reconstructed image. Although the scheme is able to find a better reconstruction of certain details than the FD-UNet, the final image quality is worse than the one obtained with DAR using LBP+FD-UNet as initial reconstruction method.

\begin{figure}
  \centering
  \includegraphics[width=12cm]{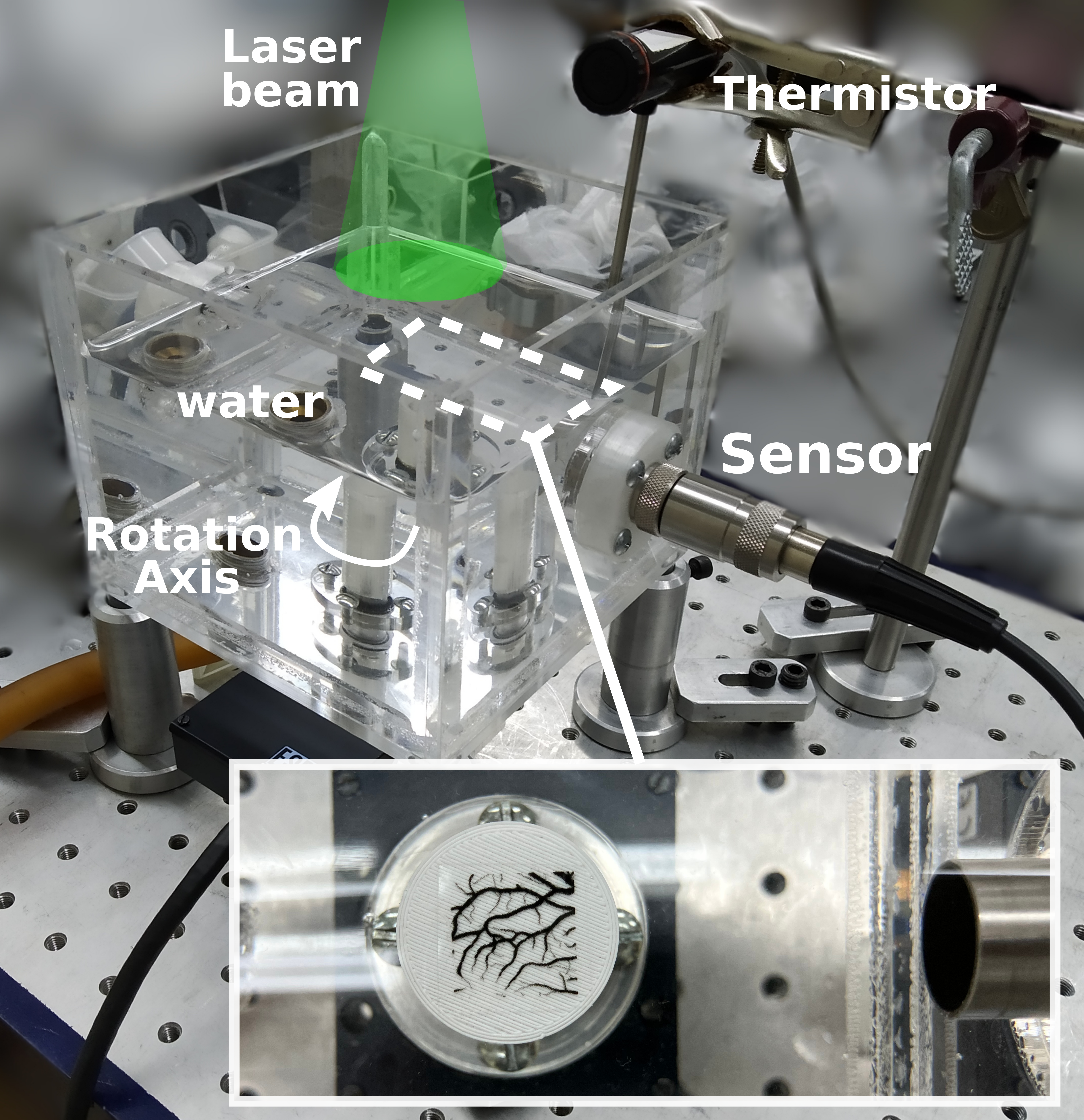}
    \caption{Photograph of the experimental setup and ground-truth image printed on a transparent film embedded in agarose gel. The insert shows the sample irradiated by the laser and the sensor tip immersed in water.}
\label{fig:medsetup}
\end{figure}

\begin{figure}
  \centering
  \includegraphics[width=16cm]{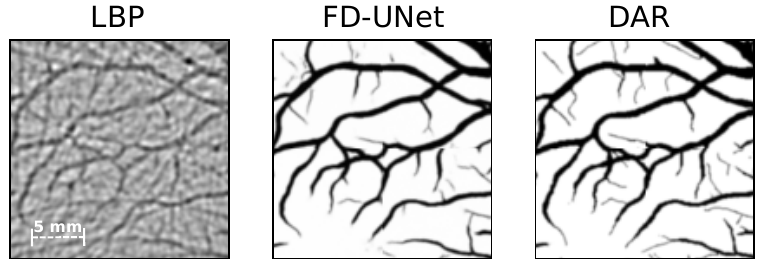}
    \caption{Reconstructed images for LBP, FD-UNet and DAR with NIS 25. PSNR is 9.97, 20.34 and 23.90 dB respectively.}
\label{fig:medres}
\end{figure}

\subsection{Experimental results}

With the goal of testing the performance of our method under experimental conditions, we assembled a two-dimensional setup as the one in Fig. \ref{fig:setup}. In Fig. \ref{fig:medsetup} a photograph of the experimental setup is presented. In this case, the sample consists of an ink pattern laser (artificial vein image not present in the training set used to optimize the parameters in the FD-UNet and DAR schemes) printed on a transparent film embedded in agarose gel (see inset in Fig. \ref{fig:medsetup}) following the steps described in \cite{Insabella_2020}. In order to simplify the experimental setup, we measured the acoustic pressure with only one detector (Olympus V306-SU) and used a stepped motor to rotate (Newport PR50C) the sample around its vertical axis in $36$ different positions. At each position, a laser illuminated the sample and the detector registered the pressure generated. In this way, we replicate the case in which $36$ detectors are simultaneously sampled, but with a considerable less complex acquisition system \cite{tian2020}. The sensor and the sample were immersed in a vessel filled with deionized water with a volume of $14\times14\times10 \text{ cm}^3$. The water temperature was measured with a calibrated thermistor to determine the speed of sound ($\sim1490 \text{ m/s}$). A Nd:YAG laser with second harmonic generation (Continuum Minilite I, 532 nm), 5 ns pulse duration, 10 Hz repetition rate and pulse energy less than 10 mJ, was the light source. A diverging lens adapted the diameter of the laser beam to a size larger than the sample, trying to achieve homogeneous illumination and a laser fluence of $2.5 \,\text{mJ/cm}^2$.  The distance between the sensor and the center of the rotating sample was $\sim44 \text{ mm}$. The sensor output was amplified with a transimpedance amplifier (EG$\&$G Optoelectronics Judson PA-400), digitized by an oscilloscope (Tektronix TDS 2024) and processed on a personal computer. The PA signals were not averaged. A pyroelectric detector (Coherent J-10MB-LE) measured the laser pulse energy. 

In Fig. \ref{fig:medres} we see the reconstructed images from the LBP, FD-UNet and DAR method. In this last case, we considered a NIS of 25. We can see that the reconstruction given by the DAR is visually closer to the ground-truth image than the corresponding reconstruction for LBP and FD-UNet. DAR is able to recover some finer details that the other methods are not able to do. This is also confirmed by the value of PSNR obtained by the three methods being $9.97 \text{ dB}$ for the LBP, $20.34 \text{ dB}$ for FD-UNet and $23.90 \text{ dB}$ for DAR. We performed further measurements of other artificial vein images printed on a transparent film embedded in agarose gel. The results were qualitatively the same as those shown in Fig. \ref{fig:medres}. The average and standard deviation values of PSNR obtained over the set of such images\footnote{We used a sample set of 8 different vein images.} were: $(23.30\pm 2.56)\text{ dB}$ for DAR with NIS 25, $(20.04\pm 2.43)\text{ dB}$ for FD-UNet and $(9.29\pm 0.62)\text{ dB}$ for LBP. As can be seen, the results are in line with those reported in Fig. \ref{fig:medres}. 

\subsection{Discussion}
\label{sec:discussion}

The use of the reverse diffusion step during inference can be thought of as an image enhancement stage by which the initial reconstructed image is improved. Our numerical and experimental results show that the improvement can be quantitative and qualitatively appreciated. Our proposal takes an initial reconstructed image using LBP+FD-UNet and uses it as conditional information to a generative reverse diffusion architecture to deliver an enhanced image. It was shown how some finer details not visible in the initial reconstructed image can be recovered by this process. Some additional remarks can be summarized as follows:
\begin{enumerate}
    \item Although there are obvious gains in terms of final reconstruction image quality they are not for free. In the first place, additional computational resources are needed both for training and for the inference stage. Although our proposal architecture can be trained independently of the initial reconstruction method, appropriate training data and computational resources are needed. This is especially true for the training stage, which is expensive in terms of some resources like memory GPU. This is a consequence of the fact that useful diffusion models usually have a large number of trainable parameters. In addition, this number dramatically increases with the required size of the reconstructed images. Similarly, for inference, additional computations are needed with respect to the initial reconstruction method. Luckily, as explained in the previous sections, the number of NIS can be significantly reduced, which also has direct influence in the actual processing time required to transform a measured sinogram into an image and as consequence in a potential real-time implementation.  
    \item It is worth noticing that given the sequential flow in the architecture proposed (LBP$\rightarrow$ FD-UNet $\rightarrow$ CIP+Reverse Diffusion) a conditional implementation of CIP+Reverse Diffusion can be implemented at inference time. For example, after inspection of the initial image reconstruction, specialized personnel in medical imaging could conclude that no further image enhancement is required. In that case, the CIP+Reverse Diffusion stage would not be implemented. However, if it is concluded that further inspection of finer details is needed, CIP+Reverse Diffusion can be activated. In this way a better use of computational resources can be implemented. 
    \item The proper design and implementation of the CIP block could have a major weight in the final performance of the Reverse Diffusion procedure. In our proposal, we have not used a principled design for the CIP. We worked with a standard autoenconder structure for its implementation and training without any consideration of the specific domain which the images to be reconstructed belongs to. It is possible that further improvements could be obtained if some image domain specifications (e.g. anatomical information) can be used to guide the CIP design. 
    \item From the previous remarks, it is clear that one of the major shortcomings of the proposed architecture is its demand on computational resources. Nevertheless, the proposed scheme shows potential benefits that can be further exploited with larger computational capabilities. For example, we could explore the design of a CIP working directly in the image space and not in a latent or dimensional reduced one, no need of patching for the diffusion process, etc.
    \item Although all blocks in our proposal can be optimized separately simplifying the search for optimal parameters, if some modification on one of them is introduced, this would possibly require the re-training of all other blocks. For example, the initial reconstruction block can be changed to other potential methods, e.g. changing the LBP for Delay and Sum (DAS) algorithm. However, this could require the re-training of the FD-UNet, the CIP and the Reverse Diffusion architecture. In this sense, the proposed architecture is not fully modular.
\end{enumerate}

\section{Conclusions}
\label{sec:conclu}
In this work, we presented a diffusion assisted method to increase the quality of reconstructed OAT images. Our method consists in a first reconstruction stage in which an initial LBP reconstruction plus a FD-UNet architecture deliver an initial image. This image is in general of good and acceptable quality. However, in order to obtain finer details, a second stage, in which a diffusion model is applied, is used to further increase image quality. This diffusion stage takes the initial image, delivered by the LBP in combination with the FD-UNet as conditional information, and generates a sample of a refined image which is a better approximation to the ground-truth image. The proposal presents good performance metrics (e.g, PSNR and SSIM) and in general is able to resolve finer details and structures that are not recoverable without using the diffusion method. The complexity added by the diffusion stage (after the model parameters are trained) do not significantly increase the processing time elapsed between sinogram acquisition and final image delivery, which is important for potential use in real-time OAT systems.

Some future lines of research could explore the use of latent diffusion, for which the tuning of the autoencoder used for mapping between image space to latent space and vice-versa should be studied and implemented. As the conditional information preprocessing is a critical block, because it is responsible for biasing the diffusion to the appropriate manifold to which the ground-truth image belongs to, some principled guidelines on designing this block could be explored. Finally, other approaches of probabilistic diffusions applied to inverse problems, as the ones in \cite{Cui_Cao_Cheng_Jia_Zheng_Liang_Zhu_2023,Aali_Arvinte_Kumar_Tamir_2023, Feng_Smith_Rubinstein_Chang_Bouman_Freeman_2023,Song_Shen_Xing_Ermon_2022}, could be studied considering the specifics of the OAT problem.

\section* {Acknowledgments}
This work was supported by the University of Buenos Aires (grant UBACYT 20020190100032BA), CONICET (grant PIP 11220200101826CO) and the ANPCyT (grant PICT 2020-01336).


\section* {Data Availability Statement}
The data that support the findings of this study are available from the corresponding author upon reasonable request.


\bibliographystyle{apalike} 
\bibliography{references}  

\end{document}